\documentclass{article}
\newcommand{\cs}[3]{{{#3} \brace {#1 #2}}}

\newcommand{\edf}{\ {\mathop{=}\limits^{\rm def}}\ }

\newcommand{\al}{\alpha}

\begin{document}

\title{ Motion of Spinning Density and Spinning Deviation Density in Riemannian Geometry}         
\date{}          
\maketitle
\begin{center}
{\author {Magd E. Kahil{\footnote{ Faculty of Engineering, Modern Sciences and Arts University, Giza, Egypt  \\
e.mail: mkahil@msa.edu.eg}}}, {Samah A. Ammar} \footnote{Women's College for arts,Science and education, Ain Shams University, Cairo, Egypt \\ e.mail: Samah.Ammar@women.asu.edu.eg} and {Shymaa A. Refaey} \footnote{Women's College for arts,Science and education, Ain Shams University, Cairo, Egypt \\ e.mail:Shymaa-Refaey@women.asu.edu.eg  } }
\end{center}
\abstract {Equations of motion of spinning density for extended objects and corresponding deviation equations are derived. The problem of motion for a variable mass to a spinning extended object is obtained. Spinning fluids may be considered as a special case to express the motion of spinning density for extended objects. Meanwhile, the spinning density tensor can be expressed in terms of
tetrad formalism of General Relativity to be regarded as a gauge theory of gravity. Equations of spinning and spinning deviation density tensors have been derived using a specific type of Bazanski
Lagrangian.}
\section{Introduction}
Spinning motion is regarded as one of the actual features of the characteristic behavior for objects in nature, which led many authors to focus on the cause of the spinning process. Would be eligible to include its internal properties or discard them as a step of simplification? From this perspective, it is vital to begin with equations of motion of spinning Mathisson-Papapetrou \cite{Papapetrou1951}
\begin{equation}
\frac{D P^{\alpha}}{Ds} = \frac{1}{2} R^{\alpha}_{.~\beta \gamma \delta}S^{\gamma\delta} U^{\beta}
\end{equation}
   where $P^{\alpha}$ is the momentum of the particle
    $$P^{\alpha}= (mU^{\alpha}+ U_{\mu} \frac{D S^{\alpha\mu} }{Ds}),$$  $R^{\alpha}_{.~\beta \gamma \delta}$ is the Riemannian curvature, $U^{\alpha}= \frac{dx^\alpha}{ds}$ is the unit tangent vector, $s$ is a parameter varying along the curve and  $S^{\gamma\delta}$ is the spin tensor. A spinning object with precession (Gyroscopic motion) can be described using the following equation
\begin{equation}
\frac{D S^{\mu \nu}}{Ds} = P^{\mu}U^{\nu} -P^{\nu}U^{\mu}.
\end{equation}
If $P^{\alpha}= mU^{\alpha},$ then equation (1.1) and (1.2) become the Papapetrou equation for short!
\begin{equation}
\frac{D U^{\alpha}}{Ds} = \frac{1}{2m} R^{\alpha}_{~\beta \gamma \delta}S^{\gamma\delta} U^{\beta},
\end{equation}
and
\begin{equation}
\frac{D S^{\mu \nu}}{Ds} = 0,
\end{equation}
provided that \cite{Kahil2018b}
\begin{equation}
S^{\mu \nu} = \bar{s} ( U^{\mu} \Psi^{\nu} - U^{\nu} \Psi^{\mu}  )
\end{equation}
where, $\bar{s}$ spin magnitude, $U^{\alpha}$ is four vector velocity and $\Psi^{\alpha}$ is a geodesic deviation vector.
Equation (1.3) can be obtained from geodesic equations \cite{Bazanski1989}
\begin{equation}
\frac{{DU}^{\alpha}}{D s} =0
\end{equation} and the geodesic deviation equations
\begin{equation} \frac{D^2 \Psi^{\alpha}}{D s^2} = R^{\alpha}_{~\mu \nu \rho} U^{\mu} U^{\nu} \Psi^{\rho}.  \end{equation}
If we apply the following transformation of paths for different parameters \cite{Kahil2018b}
\begin{equation}
\frac{ dx^{\alpha}}{ds} = \frac{dx^{\alpha}}{d \tau} + \beta \frac{D \Psi^{\alpha}}{D \tau}
\end{equation}
where $\tau$ is a parameter describing another trajectory and ${\beta}$ is an arbitrary parameter.\\ Thus,  by operating covariant derivative with respect to the parameter $s$ on both sides and taking in consideration $\frac{D\tau}{Ds}=1$,  to get
\begin{equation}
\frac{D U^{\alpha}}{Ds} = \frac{D U^{\alpha}}{D\tau} + \beta \frac{D^{2} \Psi^{\alpha}}{D \tau^{2}}
\end{equation}
Using geodesic equations (1.6)  and geodesic deviation equations (1.7) as well Equation (1.3) one can obtain Equation (1.1), while if one consider Frenkel condition
$$ S_{\mu \nu}U^{\mu} =0,$$ to be covariant differentiated on both sides  and after some  manipulations one can get Equation(1.4).

 Thus, due to the extension of spin tensor from pole-dipole moments to multi-pole moments for extended objects, this may lead to examining its corresponding propagation equation \cite{Yasskin1980}. Such an equation may be obtained by means of introducing the spin tensor density$S^{\alpha \beta \gamma}$ as a third-order skew-symmetric tensor viable to describe extended objects. These equations play a vital role in astrophysics and early cosmology to become a good candidate for describing a spinning fluid and also for describing the status of the accretion disc orbiting a compact gravitational field as in AGN \cite{Kleidis2000}. Also, it contributes to understanding the problem of motion of quark-gluon heavy ion collisions in early Universe \cite{Cao2022}.

In our present work, we are going to derive equations of spinning density tensor and spin deviation density tensor of different cases as described in Riemannian geometry in GR.
Accordingly, the spin density tensor is related to thermodynamic variables \cite{Chrohok2002}.
\begin{equation}
Td\hat{s}= d E + p d(\frac{1}{\rho})-\frac{1}{2}\omega_{\mu \nu}ds^{\mu \nu}
\end{equation}
where $T$ is the temperature, $\hat{s}$ is the entropy, $ E$ is energy density, $\omega_{\mu \nu}$ is the spin angular velocity, and $s^{\mu \nu}$ spin density. Thus, in self-consistent theories described the entropy become conserved, such that
\begin{equation}
\frac{d \hat s}{ d s} \approx1.
\end{equation}
Thus, the first law of thermodynamics becomes
\begin{equation}
\frac{d E}{ds} + p \frac{d{\rho}^{-1}}{ds} - \frac{1}{2}\omega_{\mu \nu} \frac{ds^{\mu \nu}}{ds} =1.
\end{equation}
Meanwhile, it is worth mentioning that in the presence of strong spinning a viscosity then should be included to define the spin density tensor \cite{Becattanini2019}. It is well known that spinning fluids are dominating properties of nature, which may be found to describe the problem of motion of particles in an accretion disc as a Gyrodynamics fluid. This is the counterpart of the Papapetrou equation \cite{Mosheni2008}. Owing to the spin density tensor, an interaction between spinning motion and thermodynamic variables may be found in equation (1.12) \cite{Ray1982}.

Thus, it is well known that $S^{\alpha \beta \gamma}$ is a third-order tensor, viable to define extended objects.
Nevertheless, from a progenitor case, the spin density tensor is bounded to be skew-symmetric in the last two indices, it comes to arise for expressing a spinning fluid element as a confided case of an extended object. Yet, one may find out that the Weyssenhoff tensor \cite{Cao2022} is the most eligible candidate to express a spin fluid element i.e.
\begin{equation}
{S}^{\rho \mu \nu} = S^{\mu\nu} U^{\rho}.
\end{equation}
 Thus, differentiating both  sides of (1.13) by the covariant derivative to get
\begin{equation}
\frac{D{S}^{\rho \mu \nu}}{Ds} = \frac{D S^{\mu \nu}}{Ds} U^{\rho} + \frac{D U^{\rho}}{Ds} S^{\mu \nu}.
\end{equation}
Now, one uses the following Lagrangian \cite{Kahil2018a}
\begin{equation}
L= g_{\mu \nu} U^{\mu} \frac{D \Psi^{\nu}}{Ds} + S_{\mu \nu}\frac{D \Psi^{\mu \nu}}{Ds}.
\end{equation}
By taking the variation with respect to the deviation vector $\Psi^{\rho}$
one gets (1.6),
$$\frac{D U^{\rho}}{Ds} =0.$$
Also, taking the variation with respect to the spinning deviation tensor $\Psi^{\rho \delta}$ to get (1.4)
\begin{equation}
\frac{D S^{\rho \delta}}{Ds} =0.
\end{equation}
Consequently, substituting from (1.6) and (1.16) into (1.14), one obtain
\begin{equation}
\frac{D S^{\rho \delta \sigma}}{Ds} =0,
\end{equation}
which indicates the equation of spin density tensor.
Accordingly, we may obtain equations analogously by means of its corresponding Bazanski Lagrangian stemmed from its original formalism \cite{Bazanski1989} and its modification in GR \cite{Kahil2006} to become
\begin{equation}
L=  S_{\alpha\mu \nu}\frac{D \Psi^{\alpha \mu \nu}}{Ds},
\end{equation}
such that by taking the variation with respect to $\Psi^{\rho \delta \lambda}$,
\begin{equation}
\frac{D S^{\rho \delta \lambda}}{Ds} =0.
\end{equation}
If we apply the commutation relation in such that
\begin{equation}
(S^{\rho \mu \nu}_{~~~;\alpha \beta }- S^{\rho \mu \nu}_{~~~;\beta \alpha }) U^{\alpha}\Psi^{\beta} = S^{\sigma[ \mu \nu}R^{\rho]}_{~\sigma \alpha \beta} U^{\alpha}\Psi^{\beta},
\end{equation}
and
\begin{equation}
S^{\rho \mu \nu}_{~~~; \delta } \Psi^{\delta} =  \Psi^{\rho \mu \nu}_{~~~; \delta } U^{\delta},
\end{equation}
to obtain its corresponding spin density deviation tensor equation
\begin{equation}
\frac{D^{2} \psi^{\rho \delta \lambda}}{Ds^{2}} =S^{\sigma [\delta \lambda }R^{\rho ]}_{~\sigma \alpha \beta} U^{\alpha} \Psi^{\beta}.
\end{equation}
The importance of spin density deviation tensor is to examine the stability conditions of an accretion disc orbiting very strong in gravitational field such as SgrA* \cite{Kahil2015}.

Accordingly, the paper is organized as follows. Section $2$, discusses spinning density tensor and spinning density deviation tensor equations: Papapetrou-like equations. Section $3$ displays the spinning density tensor and spinning density deviation tensor equations: a variable mass. Section $4$ emphasizes the concept of spinning density tensor and spinning density deviation tensor equations: a spinning fluid. Section $5$ performs spinning and spinning density deviation equations for a gauge theory of gravity. Section $6$ presents the conclusion and future work.
\section{Spinning Density Tensor and Spinning Density Deviation Tensor Equations: Papapetrou-Like Equations}

In this section, we are going to examine a massive density spin tensor able to describe an orbiting extended object for a compact object. Accordingly, the Weyssenhoff spin vector may be amended to be expressed as follows:

\begin{equation}
\bar{S}^{\rho \mu \nu} = S^{\mu\nu} P^{\rho},
\end{equation}
where $P^{\rho}$ is the momentum in which it is relating to $\bar{S}^{\rho \mu \nu}$ in the following sense
\begin{equation}
\bar{S}^{\rho \mu \nu} = S^{\mu\nu} (m U^{\rho}+ U_{\delta}\frac{D S^{\rho \delta}}{Ds} ),
\end{equation}
i.e.
\begin{equation}
\bar{S}^{\rho \mu \nu} = S^{\mu\nu} (m U^{\rho}+ U_{\delta}(P^{\rho} U^{\delta}- P^{\delta} U^{\rho} ) ).
\end{equation}
 Differentiating both sides by covariant derivative for (2.23) to get
\begin{equation}
\frac{D \bar{S}^{\rho \mu \nu}}{Ds} = \frac{D S^{\mu \nu}}{Ds} P^{\rho} + \frac{D P^{\rho}}{Ds} S^{\mu \nu}.
\end{equation}
We suggest the equivalent Bazanski Lagrangian to be:
\begin{equation}
 L=  \bar{S}_{\rho \mu \nu} \frac{D \bar{\Psi}^{\rho \mu \nu}}{{D}{s}} + {f}_{\rho \mu \nu}\bar{\Psi}^{\rho \mu \nu}.
 \end{equation}
Thus, by operating the Euler-Lagrange equations with respect to the deviation tensor $\bar\Psi^{\rho \mu \nu}$, which have the form
 \begin{equation}\frac{d}{ds}\frac{\partial L} {\partial\dot{\bar{\Psi}}^{\rho \mu \nu}} -\frac{\partial L}{\partial \bar\Psi^{\rho \mu \nu} } =0,\end{equation}
 we get
\begin{equation}
\frac{D \bar{S}^{\rho \mu \nu}}{Ds}= f^{\rho \mu \nu},
\end{equation}
i.e. $$f^{\rho \mu \nu}=\frac{D S^{\mu \nu}}{Ds} P^{\rho} + \frac{D P^{\rho}}{Ds} S^{\mu \nu}.$$
Consequently,
$$f^{\rho \mu \nu}=(P^{\mu} U^{\nu} - P^{\nu}U^{\mu}) P^{\rho} + \frac{1}{2} R^{\rho}_{~\alpha\sigma\beta} S^{\alpha\sigma\beta}  S^{\mu \nu},$$
where  $$ f^{\mu} = \frac{1}{2} {R}^{\mu}_{~\nu \rho \delta} S^{\nu\rho \delta},$$ is regarded as a spin force,  and $$ M^{\mu \nu} =P^{\mu}U^{\nu}- P^{\nu}U^{\mu}.$$
Consequently, its corresponding spin density deviation tensor equation can be obtained by applying in a similar way the commutation relations as given in (1.21) and (1.22) as follows:
\begin{equation}
\frac{{D}^{2}\bar{\Psi}^{\rho \mu \nu}}{Ds^{2}}=  \bar{S}^{\delta[ \mu \nu }{{R}}^{\rho ]}_{~ \delta \alpha \beta} U^{\alpha} {\Psi}^{\beta}+ f^{\rho \mu \nu}_{~~~{;} \delta}{\Psi}^{\delta}.
 \end{equation}
\section{Spinning Density Tensor and Spinning Density  Deviation Tensor Equations: A Variable Mass }
If we consider a massive spin density tensor  whose mass is not constant but function of the parameter $(s)$ in which its corresponding Weyssenhoff tensor becomes as follows
\begin{equation}
\hat{S}^{\rho \mu \nu} = m(s)U^{\rho} S^{\mu \nu}.
\end{equation}
Differentiating
\begin{equation}\frac{D\hat{S}^{\rho \mu \nu}}{Ds} = \frac{D(m(s)U^{\rho})}{Ds} S^{\mu \nu} + m(s)U^{\rho}\frac{DS^{\mu \nu}}{Ds}.
\end{equation}
Thus we can suggest a Lagrangian obtained for spinning variable mass object
\begin{equation}
L= m(s)g_{\mu \nu}U^{\mu} \frac{D \Psi^{\nu}}{Ds} + ( m(s)_{,\rho} + \frac{1}{2} R_{\rho \alpha \beta \gamma} S^{\alpha \beta \gamma}){\Psi}^{\rho} + S_{\alpha \beta} \frac{D \hat{\Psi}^{\alpha \beta}}{Ds}+ m(s)_{,\rho}\Psi^{\rho}.
\end{equation}
where m(s) is a mass function  \cite{Kahil2019}. By applying the variation \begin{equation}\frac{d}{ds}\frac{\partial L} {\partial\dot{\Psi}^{\delta}} -\frac{\partial L}{\partial \Psi^{\delta} } =0,\end{equation} to the above Lagrangian, we get
\begin{equation}
\frac{D U^{\delta}}{Ds} = \frac{m(s)_{,\rho}}{m(s)}(g^{\delta \rho}- U^{\delta} U^{\rho}) + \frac{1}{2m(s)} R^{\delta}_{~\rho \alpha \beta} S^{\rho\alpha \beta}.
\end{equation}
Also, in a similar way the Euler Lagrange equation $$\frac{d}{ds}\frac{\partial L} {\dot{\hat{\Psi}}^{\delta \sigma}} -\frac{\partial L}{\partial \hat{\Psi}^{\delta \sigma} } =0.$$
Then by taking the variation with respect to $\hat{\Psi}^{\delta \sigma}$, we obtain
\begin{equation}
\frac{D\hat{S}^{\delta \sigma}}{Ds}=0.
\end{equation}
Thus the spinning density tensor with a variable mass may be expressed in the following way:
\begin{equation}
\frac{D \hat{S}^{\rho \mu \nu}}{Ds} = ( \frac{m(s)_{,\sigma}}{m(s)}(g^{\sigma \rho}- U^{\sigma} U^{\rho}) + \frac{1}{2m(s)} R^{\rho}_{~ \lambda \alpha \beta} S^{\lambda\alpha \beta} ) S^{\mu \nu}.
\end{equation}
Accordingly, its corresponding Bazanski Lagrangian may be expressed as
\begin{equation}
 L=  \hat{S}_{\rho \mu \nu} \frac{D \hat{\Psi}^{\rho \mu \nu}}{{D}{s}} + \hat{f}_{\rho \mu \nu}{\hat{\Psi}}^{\rho \mu \nu} ,
\end{equation}
where $ \hat{f}^{\rho \mu \nu}= ( \frac{m(s)_{,\sigma}}{m(s)}(g^{\sigma \rho}- U^{\sigma} U^{\rho}) + \frac{1}{2m(s)} R^{\rho}_{~ \lambda \alpha \beta} S^{\lambda\alpha \beta}) S^{\mu \nu}. $
By taking the variation with respect to $\hat{\Psi}^{\rho \mu\nu}$, we obtain
\begin{equation}
\frac{D \hat{S}^{\rho \mu \nu}}{Ds}= \hat{f}^{\rho \mu \nu} .
\end{equation}
As we follow the same procedure as given in equation (1.21), we obtain the deviation spinning density equation to become
\begin{equation}
\frac{D^{2} \hat{\Psi}^{\rho \mu \nu}}{Ds^{2}}= (\hat{f}^{\rho \mu \nu})_{;\delta} \Psi^{\delta} + S^{\xi[\mu \nu}R^{\rho]}_{~~\xi \alpha \beta}U^{\alpha}\Psi^{\beta}.
\end{equation}
This equation may be applied in studying the stability of a variable spinning disk which may work to explain the effect of mass excess in a region orbiting a compact object. Such an illustration may give rise to examining the effect of dark matter in the accretion disk of a compact object.
\section{Spinning Density Tensor and Spinning Density  Deviation Tensor Equations: A Spinning Fluid }
We are going to suggest that the relation between a variable mass and a spinning fluid in the following way,
\begin{equation}
\tilde{S}^{\rho \mu \nu} = (p + \rho)(s)U^{\rho} S^{\mu \nu},
\end{equation}
where,
~~~~~~~$m(s) = (p + \rho)$.\\ Accordingly, the pressure is turning to be only parameter of $ s $, while the density is becoming constant. Owing to this suggestion the spin fluid behaves like a spinning variable mass, such that:
\begin{equation}
\frac{d m(s)}{ dx^{p}} = \frac{d p (s)}{dx^{p}}.
\end{equation}
Such an equivalence may give rise to suggest the following Lagrangian,
\begin{equation}
L= (p+\rho)(s)g_{\mu \nu}U^{\mu} \frac{D \Psi^{\nu}}{Ds} + (P_{,\rho} + \frac{1}{2} R_{\rho \alpha \beta \gamma} S^{\beta \gamma} U^{\alpha} )\Psi^{\rho} + S_{\alpha \beta} \frac{D \tilde{\Psi}^{\alpha \beta}}{Ds}.
\end{equation}
Such that, by taking the variation with respect to $\Psi^{\delta}$ as given by (3.34), we get
\begin{equation}
\frac{D U^{\delta}}{Ds} = \frac{p_{,\rho}}{(p+\rho)}(g^{\delta \rho}- U^{\delta} U^{\rho}) + \frac{1}{2(p+\rho)} R^{\delta}_{\rho \alpha \beta} S^{\rho\alpha \beta}.
\end{equation}
Also, by taking the variation with respect to $\tilde{\Psi}^{\rho \delta}$, which is given by
\begin{equation}\frac{d}{ds}\frac{\partial L} {\partial{\dot{\tilde{\Psi}}^{\rho \delta}}} -\frac{\partial L}{\partial {\tilde{\Psi}}^{\rho \delta} } =0,\end{equation}we obtain
\begin{equation}
\frac{D\hat{S}^{\rho \delta}}{Ds}=0.
\end{equation}
Equation of a spinning fluid can be expressed in the following way:
\begin{equation}
\frac{D\tilde{S}^{\rho \mu \nu}}{Ds} = ( \frac{ p_{,\sigma}}{(p+\rho)}(g^{\sigma \rho}- U^{\sigma} U^{\rho}) + \frac{1}{2(p+\rho)} R^{\rho}_{~ \lambda \alpha \beta} S^{\lambda\alpha \beta}) S^{\mu \nu}.
\end{equation}
Accordingly, its corresponding Bazanski Lagrangian may be expressed as
\begin{equation}
 L=  \tilde{S}_{\rho \mu \nu} \frac{D \tilde{\Psi}^{\rho \mu \nu}}{{D}{s}} + \tilde{f}_{\rho \mu \nu}{\tilde{\Psi}}^{\rho \mu \nu} ,
\end{equation}
where, $\tilde{\Psi}^{\rho \mu \nu}$ is the spin deviation  tensor of $\tilde{S}_{\rho \mu \nu}$
and $ \tilde{f}^{\rho \mu \nu}= ( \frac{p_{,\sigma}}{p+\rho}(g^{\sigma \rho}- U^{\sigma} U^{\rho}) + \frac{1}{2(p+\rho)} R^{\rho}_{~\lambda \alpha \beta} S^{\lambda\alpha \beta} ) S^{\mu \nu}. $
By taking the variation with respect to $\tilde{\Psi}^{\rho \mu \nu}$, which have the form $$\frac{d}{ds}\frac{\partial L} {\partial\dot{\tilde{\Psi}}^{\rho \mu \nu}} -\frac{\partial L}{\partial \tilde{\Psi}^{\rho \mu \nu} } =0,$$ we obtain
\begin{equation}
\frac{D \tilde{S}^{\rho \mu \nu}}{Ds}= \tilde{f}^{\rho \mu \nu}.
\end{equation}
By following the same procedure given by (1.21) and (1.22), we get 
\begin{equation}
\frac{D^{2} \tilde{\Psi}^{\rho \mu \nu}}{Ds^{2}}= (\tilde{f}^{\rho \mu \nu})_{;\delta} \Psi^{\delta} + S^{\xi[\mu \nu}R^{\rho]}_{~~\xi \alpha \beta}U^{\alpha}\Psi^{\beta}.
\end{equation}
\subsection{Modified Forms of Spin Density}
  In this part, we suggest a modified form of spin density tensor in Riemannian geometry to be; \\
 (a)\underline{ $P^{\alpha}= mU^{\alpha}$}
\begin{equation}
  S^{\alpha \beta \gamma} = \frac{1}{3!} (S^{\beta \gamma} U^{\alpha} + S^{\gamma \alpha} U^{\beta}+ S^{\alpha \beta} U^{\gamma}).
  \end{equation}

  Differentiating both sides using covariant derivative,
\begin{equation}
  \frac{D{S^{\alpha \beta \gamma}}}{Ds} = \frac{1}{3!} (\frac{D S^{\beta \gamma}}{Ds} U^{\alpha} + S^{\beta \gamma} \frac{D U^{\alpha}}{Ds}+ \frac{S^{\gamma \alpha}}{Ds} U^{\beta}+S^{\gamma \alpha} \frac{D U^{\beta}}{Ds}+
  \frac{D S^{\alpha \beta}}{Ds}U^{\gamma}+ S^{\alpha \beta} \frac{D U^{\gamma}}{Ds} ).
  \end{equation}
  (i) For $ \frac{DU^{\alpha}}{Ds} =0 $ and $ \frac{DS^{\alpha\beta}}{Ds} =0$ and substituting in (4.51), we obtain
\begin{equation}
  \frac{DS^{\alpha \beta \gamma}}{Ds} =0.
  \end{equation}
  (ii) For $\frac{DU^{\alpha}}{Ds}= \frac{1}{2m}R^{\alpha}_{~\beta \gamma \delta} S^{\gamma \delta} U^{\beta} $  and $ \frac{DS^{\alpha\beta}}{Ds} =0$, (4.51) can be rewritten as
\begin{equation}
  \frac{DS^{\alpha \beta \gamma}}{Ds} =\frac{1}{3!}(\frac{1}{2m}R^{\alpha}_{~\mu \nu \rho } S^{\nu \rho} U^{\mu} S^{\beta\gamma}+ \frac{1}{2m}R^{\beta}_{~\mu \nu \rho} S^{\nu \rho} U^{\mu} S^{\gamma \alpha}+\frac{1}{2m}R^{\gamma}_{~\mu \nu \rho} S^{\nu \rho} U^{\mu}S^{\alpha \beta}),
  \end{equation}
  i.e.
 \begin{equation}
  \frac{DS^{\alpha \beta \gamma}}{Ds} =\frac{1}{3!}( \frac{1}{2}(R^{\alpha}_{~\mu \nu \rho } S^{\beta\gamma}+ R^{\beta}_{~\mu \nu \rho} S^{\gamma \alpha}+R^{\gamma}_{~\mu \nu \rho} S^{\alpha \beta}) S^{\mu\nu \rho}),
  \end{equation}
  to become
\begin{equation}
  \frac{DS^{\alpha \beta \gamma}}{Ds} =\frac{1}{2m}(R^{(\alpha}_{~~\mu \nu \rho } S^{\beta\gamma)}) S^{\mu\nu \rho},
  \end{equation}
  (b)\underline{$P^{\alpha}\neq mU^{\alpha}$}
\begin{equation}
  \bar{S}^{\alpha \beta \gamma} = \frac{1}{3!} (S^{\beta \gamma} P^{\alpha} + S^{\gamma \alpha} P^{\beta}+ S^{\alpha \beta} P^{\gamma}).
  \end{equation}
  Differentiating both sides using covariant derivative,
\begin{equation}
  \frac{D{\bar{S}^{\alpha \beta \gamma}}}{Ds} = \frac{1}{3!} (\frac{D S^{\beta \gamma}}{Ds} P^{\alpha} + S^{\beta \gamma} \frac{D P^{\alpha}}{Ds}+ \frac{DS^{\gamma \alpha}}{Ds} P^{\beta}+S^{\gamma \alpha} \frac{D P^{\beta}}{Ds}+
  \frac{D S^{\alpha \beta}}{Ds}P^{\gamma}+ S^{\alpha \beta} \frac{D P^{\gamma}}{Ds} ).
  \end{equation}
However, if $\frac{DP^{\alpha}}{Ds} =\frac{1}{2}R^{\alpha}_{~\beta \gamma \delta} S^{\gamma \delta}U^{\beta}$ and $\frac{DS^{\alpha \beta}}{Ds} = P^{\alpha}U^{\beta}- P^{\beta}U^{\alpha}$
  then, we obtain
  $$
  \frac{D{\bar{S}^{\alpha \beta \gamma}}}{Ds} = \frac{1}{3!} ((P^{\beta} U^{\gamma}- P^{\gamma}U^{\beta}) P^{\alpha} + \frac{1}{2}R^{\alpha}_{~\mu \nu \rho} S^{\mu \nu \rho}S^{\beta \gamma} + (P^{\gamma}U^{\alpha}-P^{\alpha}U^{\gamma})P^{\beta}
  $$
\begin{equation}
  ~~~~~~~~~+ \frac{1}{2}R^{\beta}_{~\mu\nu \rho} S^{\mu \nu \rho}S^{\gamma \alpha}+(P^{\alpha}U^{\beta}-P^{\beta}U^{\alpha})P^{\gamma}+  \frac{1}{2}R^{\gamma}_{~\mu\nu\rho}S^{\mu \nu \rho}S^{\alpha \beta} ),
  \end{equation}
  in which can be
\begin{equation}
  \frac{D{\bar{S}^{\alpha \beta \gamma}}}{Ds} = \frac{1}{12} S^{\mu\nu\sigma}(S^{\alpha\beta}R^{\gamma}_{~\mu \nu\sigma}+S^{\beta\gamma}R^{\alpha}_{~\mu \nu\sigma}+S^{\gamma\alpha}R^{\beta}_{~\mu \nu\sigma} ).
  \end{equation}
To get their corresponding deviation equation as similar as equation (4.50).
\section{Spinning and spinning density deviation equations for a gauge theory of gravity : Tetrad formalism}
\subsection{A class of Gauge theories of gravity in Riemannian geometry}
The problem of studying the microscopic structure of particles led many relativists to get a paradigm shift towards gauge theories of gravity. From this perspective, the building blocks of this type of theories stands for an analogy between the quantities used in current gauge theories and space-time. It is of interest to revisit primarily GR to be descriptive by gauge theory and their counterpart in non-Riemannian geometry \cite{Hehl1976, Hammond2002}. Yet, Collins et al (1989) described GR as a gauge theory of gravity subject to the following gauge potential vectors $e^{a}_{\mu}$ and $\omega^{ij}_{.~. \mu}$ to represent translational and rotational gauge potentials respectively. From this perspective, the problem of invariance of any quantities  must be a covariant derivative invariant under general coordinate transformation (GCT) and Local Lorentz transformation (LLT) that are expressible in terms of gauge potential of translation and rotation in the following way \cite{Collins1989}.
The equations of physics will contain derivatives of tensor fields and it is, therefore, necessary to define the covariant derivatives of tensor fields under the transformations GCT and LLT, one must define two types of connection fields to be associated with each of them. Accordingly, the Christoffel symbol $\cs{\mu}{\nu}{\alpha}$ is referred to GCT while the spin connection $\omega_{a b \mu}$ as related to LLT. \\
Accordingly ,
\begin{equation}
 D_{\mu}e^{m}_{\nu}  \edf e^{m}_{\nu , \mu} - \{^{\al}_{\nu\mu}\}e^{m}_{\alpha} + \omega_{\mu~.~n}^{~.m ~.}e^{n}_{\nu}.
\end{equation}
 Provided that
\begin{equation}
  D_{\rho}D_{\mu}e^{m}_{\nu}=  D_{\mu}D_{\rho}e^{m}_{\nu}.
  \end{equation}
 Using this concept it turns out that GR may be expressed in terms of connecting $e^{\mu}_{a}$ , $\omega_{ a b \mu }$ and $\cs{\mu}{\nu}{\alpha}$ together
\begin{equation}
 g_{\mu \nu}  \edf e^{a}_{\mu}e^{b}_{\nu} \eta_{a b},
 \end{equation}
\begin{equation}
 \{^{\al}_{\mu\nu}\} \edf \frac{1}{2} g^{\alpha \sigma}(g_{\nu \sigma , \alpha}+g_{\sigma \alpha, \nu} -g_{\alpha \nu , \sigma}  ).
 \end{equation}
 And for any arbitrary vector $A_{\mu}$
\begin{equation}
 A_{\mu ; c  d} - A_{\mu ; d c } = {{\breve{R}^{\alpha}}}_{~\mu d c} A_{\alpha},
\end{equation}
  where $ {{\breve{R}^{\alpha}}}_{~\mu d c}$ the curvature tensor may be defined, due to gauge approach, in terms of spin connection $\omega^{a}_{~b \mu}$
$$
 {{\breve{R}^{c}}}_{~d \mu \nu} \edf  \omega^{c}_{~d \nu , \mu} - \omega^{c}_{~d \mu , \nu} + \omega^{c}_{~a \mu }\omega^{a}_{~d \nu} - \omega^{c}_{~a \nu }\omega^{a}_{~d \mu}
$$
From this perspective, one can figure out the effect of both mesh indices and world indices on describing the curvature tensor.
\subsection{Spinning and Spinning Deviation equation of an object in a class of gauge theory in Riemannian geometry }
In a similar way to \cite{Collins1989}, one can find out the equations of motion for a spinning object with precession for a class of gauge theory by taking the variation with respect to $\Psi^{\mu}$ and $\Psi^{\mu \nu}$  simultaneously, for the following Lagrangian
\begin{equation}
L = g_{\mu \nu} P^{\mu} \frac{D \Psi^{\nu}}{Ds}+\frac{1}{2} {\breve{R}}_{\mu  \nu a b } S^{ ab}U^{\nu} \Psi^{\mu} + e^{a}_{\mu} e^{b}_{\nu} S_{a b} \frac{D \Psi^{\mu \nu}}{Ds} + (P_{\mu}U_{\nu}- P_{\nu}U_{\mu})\Psi^{\mu \nu},
\end{equation}
to obtain
 \begin{equation}
\frac{D P^{\alpha}}{Ds} = \frac{1}{2} {{\breve{R}^{\alpha}}}_ {~\nu a b } S^{a b}U^{\nu}.
\end{equation}
Multiply both sides by $e^{c}_{\alpha}$
 \begin{equation}
\frac{D P^{c}}{Ds} = \frac{1}{2} {{\breve{R}^{c}}}_ {~\nu a b } S^{a b}U^{\nu},
\end{equation}
and
 \begin{equation}
 \frac{ D S^{\alpha \beta}}{Ds} = ( P^{\alpha} U^{\beta}- P^{\beta}U^{\alpha}),
\end{equation}
i.e.
 \begin{equation}
 \frac{D e^{a}_{\alpha} e^{b}_{\beta} S^{a b}}{Ds} = ( P^{\alpha} U^{\beta}- P^{\beta}U^{\alpha}),
\end{equation}

 \begin{equation}
e^{a}_{\alpha} e^{b}_{\beta} \frac{D  S^{a b}}{Ds}+ S^{ab}{} \frac{D ({e^{a}_{\alpha} e^{b}_{\beta}})}{Ds}= ( P^{\alpha} U^{\beta}- P^{\beta}U^{\alpha}),
\end{equation}
multiplying both sides by $ e^{c}_{\alpha} e^{d}_{\beta} $
 \begin{equation}
e^{c}_{\alpha} e^{d}_{\beta} e^{a}_{\alpha} e^{b}_{\beta} \frac{D  S^{a b}}{Ds}+ e^{c}_{\alpha} e^{d}_{\beta}S^{ab} \frac{D ({e^{a}_{\alpha} e^{b}_{\beta}})}{Ds}= e^{c}_{\alpha} e^{d}_{\beta}( P^{\alpha} U^{\beta}- P^{\beta}U^{\alpha}),
\end{equation}
provided that
 \begin{equation}
 \frac{D ({e^{a}_{\alpha} e^{b}_{\beta}})}{Ds}=0,
\end{equation}
 consequently,
 \begin{equation}
 \frac{D  S^{c d}}{Ds}= ( P^{c} U^{d}- P^{d}U^{c}).
\end{equation}
Using the commutation relations as mentioned above, we obtain their corresponding deviation equations;
 \begin{equation}
\frac{D^2 \Psi^{c d}}{Ds^2} =  S^{l[d }{{\breve{R}^{\alpha]}}} _{~l \gamma \delta} U^{\gamma} \Psi^{\delta} + (P^{c}U^{d} - P^{d}U^{c})_{; \rho} \Psi^{\rho}.
\end{equation}
\subsection{Spinning Density and Spinning Density Deviation Equations for a class of Gauge theory}
Let us define the following spin tensor
 \begin{equation}
\bar{S}^{abc} = S^{bc} P^{a},
\end{equation}
Differentiating both sides using covariant derivative to get
 \begin{equation}
\frac{D \bar{S}^{a b c }}{Ds} = \frac{D S^{bc}}{Ds} P^{a} + \frac{D P^{a}}{Ds} S^{bc}.
\end{equation}
Substituting (5.68) and (5.74) in (5.77) to obtain
 \begin{equation}
\frac{D \bar{S}^{abc}}{Ds}=f^{abc}.
\end{equation}
where, $f^{abc}\edf p^{a}(p^{b}U^{c}-p^{c}U^{b})+\frac{1}{2} {\breve{R}^{a}}_{~\mu i j} S^{\mu i j}S^{bc}.$\\
We suggest the equivalent Bazanski Lagrangian
 \begin{equation}
 L=  \bar{S}_{abc} \frac{D \bar{\Psi}^{abc}}{{D}{s}} + f_{abc}\bar{\Psi}^{abc}.
 \end{equation}
To become after taking the variation with respect to its corresponding deviation tensor $\bar{\Psi}^{abc}$
 \begin{equation}
\frac{D \bar{S}^{abc}}{Ds}= f^{abc}.
\end{equation}
Similarly their corresponding geodesic deviation equation may be found as follows;
 \begin{equation}
\frac{{D}^{2}\bar{\Psi}^{abc}}{Ds^{2}}=  S^{d[ bc }{{\breve{R}^{a]}}}_{~d ij} S^{d i j}+ f^{abc}_{~~~{;}\gamma} \Psi^{\gamma}.
 \end{equation}
\section{Discussion and Concluding Remarks}
 The spinning density tensor may play to express a spinning fluid. From this perspective the 3rd rank skew-symmetric tensor turns to be specified to become $S^{\alpha \beta \gamma}$ skew-symmetric in the last two indices in an approach to relate the spin density tensor with the Weyssenhoff tensor which is describing a spinning fluid.
The problem of spinning density is vital to describe an extended object which can be expressible to plasma fluid and heavy ion collisions \cite{Gallegos2022}.

 In the present work, we have focused on the spinning fluid tensor and its deviation as a special case of applying a spinning density tensor. In principle, a spinning fluid tensor is a special case of spin density as it is skew-symmetric in the last two indices as expressed by Weyssenhoff tensor. This description led many authors to geometrize it by means of using Riemannian-Cartan geometry instead of the Riemannian geometry this led to a transition of GR into Einstein-Cartan theory of Gravity \cite{Ray1982,Smalley1997}. This type of geometry has employed a new role of orthonormal frames using to describe its internal composition. This gives rise to an insightful vision to examine the internal structure of a fluid element using a rotational potential vector $\Gamma^{i}_{j \mu}$ called spin connection able to describe torsion of space-time beside the translational potential vector $e^{a}_{\alpha}$. It is worth of mention  Hehl et. al, were able to include spin tensor not only of energy-momentum tensor $T_{\mu \nu}$ but also in angular momentum equations $\Omega^{\mu}_{\nu \rho}$ in terms of torsion of space-time, which is present in Poincare gauge theory (PGT)\cite{Hammond2002, Hehl1979}.

Moreover, a tetrad formalism  has been  expressed by means of orthonormal frames in order to express the spin density tensor as shown in Equation (5.76), provided  that the curvature tensor is defined in terms of spin connection terms \cite{Collins1989}. Using this type of amended curvature with symmetric affine connection, torsionless we have obtained  its corresponding spin density equation (5.78) and the spin deviation density tensor equation (5.81) as expressed in mixed way anholonomic indices besides holonomic ones.

Nevertheless,an extended work to examine the problem of motion in Clifford space \cite{Kahil2020}, a spinning density tensor may be described in Clifford spaces \cite{Castro2005} in the following way such that  $\hat{U}^{\mu}= \frac{\partial s}{\partial x^{\mu}}$ $\hat{S^{\mu \nu}} = \frac{\partial s}{\partial x^{\alpha \beta}}$ and  $\hat{S}^{\mu\nu \rho} = \frac{\partial s}{\partial x^{\alpha \beta \gamma}}$  in terms of holograhic coordinates such that $s=s(x^{\mu},x^{\mu\nu},x^{\mu\nu\rho},....)$. However, from this perspective, replacing vectors with poly-vectors leading to express torsion and curvature tensors which is viable to describe the internal structure of spinning density tensors in Riemannian geometry. This issue will be studied in detail in our future work as well.



\begin{thebibliography}{8}

 {\bibitem{Papapetrou1951} A. \texttt{Papapetrou}, "\emph{Spinning test-particles in general relativity. I}," Proc. R. Soc. Lond. A,
\textbf{209}, 248--258, (1951).}

 {\bibitem{Kahil2018b} M. E. \texttt{Kahil}, "\emph{The spinning equations of motion for objects in AP-geometry}," ADAP,
\textbf{3}, 136, (2018).}

 {\bibitem{Bazanski1989} S. L. \texttt{Bazanski}, "\emph{Hamilton--Jacobi formalism for geodesics and geodesic deviations}," J. math. phys.,
\textbf{30}, 1018--1029, (1989).}

 {\bibitem{Yasskin1980} Ph. B. \texttt{Yasskin} and W. R. \texttt{Stoeger}, "\emph{Propagation equations for test bodies with spin and rotation in theories of gravity with torsion}," Phys. Rev. D, \textbf{21}, 2081--2094, (1980).}

 {\bibitem{Kleidis2000} K. \texttt{Kleidis} and N. K. \texttt{Spyrou}, "\emph{Geodesic motions versus hydrodynamic flows in a gravitating perfect fluid: Dynamical equivalence and consequences}," Class.Quant.Grav., \textbf{17}, 2965--2982, (2000).}

{\bibitem{Cao2022} Z. \texttt{Cao}, K. \texttt{Hattori}, M. \texttt{Hongo}, Xu-Guang \texttt{Huang} and H. \texttt{Tya}, "\emph{Gyrohydrodynamics: Relativistic spinful fluid with strong vorticity}," Prog. Theor. Exp. Phys., arXiv:2205.08015}. 

 {\bibitem{Chrohok2002} Th. \texttt{Chrohok}, H. \texttt{Hermann} and G. \texttt{R\"{u}ckner}, "\emph{Spinning Fluids in Relativisitic Hydrodynamics}," Thchnicshe Meckanik,
\textbf{22}, 1, (2002).} 

 {\bibitem{Becattanini2019} F. \texttt{Becattanini}, W. \texttt{Florkowski} and E. \texttt{Speranza}, "\emph{Spin tensor and its role in non-equilibrium thermodynamics}," Phys. Lett. B, \textbf{789}, 419--425, (2019).}

 {\bibitem{Mosheni2008} M. \texttt{Mosheni}, "\emph{Spinning Fluid Cosmology}," Phys. Lett. B,
\textbf{663}, 165, (2008).} 

 {\bibitem{Ray1982} J.R. \texttt{Ray} and L. L. \texttt{Smalley}, "\emph{Spinning fluids in general relativity}," Phys. Rev. D,
\textbf{26}, 2619, (1982).}

{\bibitem{Kahil2018a} M. E. \texttt{Kahil}, "\emph{Spinning and Spinning Deviation Equations for Special Types of Gauge Theories of Gravity}," Gravit. Cosmol., \textbf{24}, 84--91, (2018).}

 {\bibitem{Kahil2006} M. E. \texttt{Kahil}, "\emph{Motion in Kaluza-Klein type theories}," J. math. phys.,
\textbf{47}, 052501, (2006).}

{\bibitem{Kahil2015} M. E. \texttt{Kahil}, "\emph{Stability of Stellar System Orbiting SGR A*}," Odessa Astro. Pub.,
\textbf{28}, 126, (2015).}

{\bibitem{Kahil2018a} M. E. \texttt{Kahil}, "\emph{Dark Matter: The Problem of Motion}," Gravit. Cosmol., \textbf{25},268--276, (2019).}


 {\bibitem{Hehl1976} F. W. \texttt{Hehl}, P. von der \texttt{Heyde}, G. D. \texttt{Kerlick} and J. M. \texttt{Nester}, "\emph{General Relativity with Spin and Torsion: Foundations and Prospects}," Rev. Mod. Phys.,
\textbf{48}, 393--416, (1976).}

 {\bibitem{Hammond2002} R. T. \texttt{Hammond}, "\emph{Torsion gravity}," Rept. Prog. Phys.,
\textbf{65}, 599--649, (2002).}

 {\bibitem{Collins1989} P. \texttt{Collins}, A. \texttt{Martin} and E. \texttt{Squires}, "\emph{Particle Physics and Cosmology},"
John Wiley and Sons, New York, (1989).}

 {\bibitem{Gallegos2022}  A. D. \texttt{Gallegos}, \texttt{G\"{u}rsory} and A. \texttt{Yarom}, "\emph{Hydrodynamics, spin currents and torsion}," ArXiv:2203.05044, (2022).}

 {\bibitem{Smalley1997} L. L. \texttt{Smalley} and J. P. \texttt{Krisch}, "\emph{String fluid dynamics in general relativity}," Class. Quant. Grav.,
\textbf{14}, (1997).}

 {\bibitem{Hehl1979} F. W. \texttt{Hehl}, "\emph{Proceedings of the 6th Course of the International School of
Cosmology and Gravitation on "Spin, Torsion and Supergravity}," P.G.\texttt{Bergamann} and V. \texttt{de Sabatta}, held at Erice,
\textbf{1}, (1979).}

 {\bibitem{Kahil2020} M. E. \texttt{Kahil}, "\emph{Motion in Clifford Space}," J. Mod. Phys.,
\textbf{11}, 1865--1873, (2020).}

 {\bibitem{Castro2005} C. \texttt{Castro} and M. \texttt{Pavsic}, "\emph{The Extended Relativity Theory in Clifford Spaces}," Prog. Phys.,
\textbf{1}, 31, (2005).}

\end{thebibliography}
 \end{document}